\documentclass[prd,aps,12pt,
nofootinbib,tightenlines,superscriptaddress,preprintnumbers]{revtex4}
\usepackage{bm}
\usepackage{amsmath}
\usepackage{yfonts}

\begin{document}

\preprint{BI-TP 2009/22}

\title{$R-$charge thermodynamical spectral  sum rule \\ in ${\cal{N}}=4$ 
Yang-Mills theory}
\author{Rudolf Baier\protect}
\affiliation{Fakult\"at f\"ur Physik, Universit\"at Bielefeld, D-33501 
Bielefeld, Germany}

\date{\today}

\begin{abstract}
A sum rule related to the $R-$current correlator at vanishing three-momentum
is derived in the ${\cal{N}}=4$ 
supersymmetric Yang-Mills field theory at infinite 't Hooft
coupling. For reference it is compared to the one in the free field theory, i.e. at the
one loop perturbative approximation.
\end{abstract}

\maketitle

\section{Introduction}
\label{sec:intro}

Relativistic heavy ion physics is confronted with phenomena which are
interpreted in terms of either a weakly interacting plasma (pQGP) or more
 recently due to experimental results produced at RHIC by the strongly-coupled
quark-gluon plasma (sQGP) \cite{Zajc:2009je}.
Transport properties are part of the main interest,
especially the hydrodynamic coefficients \cite{Romatschke:2009im},
 in order to find the proper interpretation.

\noindent
In this note the current-current correlator at finite temperature
\cite{Teaney:2006nc, Myers:2007we,CaronHuot:2006te}
is considered.
In analogy to the ``hydrodynamic''  sum rule derived in
\cite{Romatschke:2009ng}  a certain sum rule for the spectral density
in a hot prototype gauge theory of sQGP is derived, i.e. in the
${\cal{N}} = 4$ supersymmetric Yang-Mills field theory.

\section{Dispersion relation}
\label{sec:dispersion}

In the framework of linear response theory one considers the retarded
correlator \cite{Petreczky:2005nh}

\begin{equation}
\chi(t) = i \Theta(t) <[O(t), O(0)]>~,
\label{eq:e1}
\end{equation}
and its Fourier transform

\begin{equation}
\chi(\omega) = \int_{-\infty}^{+\infty} dt~ e^{i\omega t} ~\chi(t)~.
\label{eq:e2}
\end{equation}

\noindent
The spectral density is defined by

\begin{equation}
\rho(\omega) = \frac{1}{2 \pi}  \int_{-\infty}^{+\infty} dt ~e^{i\omega t}
 <[O(t), O(0)]>~,
\label{eq:e3}
\end{equation}
respectively by
\begin{equation}
<[O(t), O(0)]>= \int ~d\omega ~ e^{- i\omega t}~\rho(\omega)~.
\label{eq:e4}
\end{equation}

\noindent
Inserting (\ref{eq:e4}) into (\ref{eq:e2}) one obtains
\begin{equation}
\chi(\omega + i \epsilon) = \chi^R(\omega) =
\int_{-\infty}^{+\infty} d\omega' 
\frac{\rho(\omega')}{\omega' - \omega - i \epsilon}
= \frac{1}{\pi}\int_{-\infty}^{+\infty} d\omega' 
\frac{Im \chi(\omega')}{\omega' - \omega - i \epsilon}~.
\label{eq:e5}
\end{equation}

\noindent
This basic dispersion relation is assumed to hold at finite temperature $T$
as well as at $T=0$. Following the derivation given in \cite{Romatschke:2009ng}
consider first

\begin{equation}
\chi^R(\omega, T) - \chi^R(\omega, T=0) =
\int_{-\infty}^{+\infty} d\omega' ~
\frac{\rho(\omega', T) - \rho(\omega', T=0)}{\omega' - \omega - i \epsilon}~,
\label{eq:e6}
\end{equation}
and subtract finally the piece at $\omega \rightarrow i\infty$ - 
the integral (\ref{eq:e6}) is expected to vanish in this limit - 
to obtain \cite{Romatschke:2009ng,CaronHuot:2009ns}

\begin{eqnarray}
\chi^R(\omega, T) - \chi^R(\omega, T=0)&-&
[~\chi^R(\omega \rightarrow i \infty, T) - 
\chi^R(\omega \rightarrow i \infty, T=0)~] 
\nonumber \\
 & = & \int_{-\infty}^{+\infty} d\omega' ~
\frac{\delta \rho(\omega')}{\omega' - \omega - i \epsilon}~,
\label{eq:e7}
\end{eqnarray}
with
\begin{equation} 
\delta \rho(\omega) =\rho(\omega, T) - \rho(\omega, T=0)~.
\label{eq:e8}
\end{equation}

\noindent
Expressed in terms of the retarded Green's function $G_R(\omega) =
- \chi^R(\omega)$ (\ref{eq:e7}) can be written as

\begin{equation}
\delta G_R(\omega) - \delta G_R^{\infty} = -
\int_{-\infty}^{+\infty} d\omega' ~
\frac{\delta \rho(\omega')}{\omega' - \omega - i \epsilon}~,
\label{eq:e9}
\end{equation}
$\delta G_R(\omega) = G_R(\omega, T) - G_R(\omega, T=0)$ and
$\delta G_R^{\infty} = G_R(\omega = i \infty, T) - 
G_R(\omega = i \infty, T=0)$.

\noindent
Finally, using $\delta \rho (- \omega) = - \delta \rho (\omega)$
the result derived in \cite{Romatschke:2009ng}
\begin{equation}
\delta G_R(\omega) - \delta G_R^{\infty} = - 2
\int_{0}^{+\infty} d\omega' ~
\frac{\delta \rho(\omega')}{\omega' - \omega - i \epsilon}~,
\label{eq:e10}
\end{equation}
is obtained.

\noindent
In the following the static limit $\omega = 0$ is considered
\begin{equation}
\delta G_R(\omega = 0) - \delta G_R^{\infty} = - 2
\int_{0}^{+\infty} d\omega' ~
\frac{\delta \rho(\omega')}{\omega'}~,
\label{eq:e11}
\end{equation}
i.e. a so-called thermodynamic sum rule \cite{LeBellac}.

\section{Euclidean correlator}
\label{sec:correlator}

Let us start from the Euclidean correlator \cite{Teaney:2006nc}
\begin{equation}
G(\tau) = \int_{0}^{\infty} d \omega~ \rho(\omega)
\frac{\cosh[\omega(\tau - \beta/2)]}{\sinh(\omega \beta/2)}~,
\label{eq:ec1}
\end{equation}
and take the integral ($\beta = 1/T$)
\begin{equation}
\int_{0}^{\beta} d\tau~ G(\tau) = 
\int_{0}^{+\infty} d\omega ~
\frac{ \rho(\omega)}{\omega}~.
\label{eq:ec2}
\end{equation}
Together with the definition of the reconstructed correlator
\cite{Datta:2003ww}
\begin{equation}
G^{rec}(\tau) = \int_{0}^{\infty} d \omega~ \rho(\omega, T=0)
\frac{\cosh[\omega(\tau - \beta/2)]}{\sinh(\omega \beta/2)}~,
\label{eq:ec3}
\end{equation}
i.e. taking the spectral density at zero temperature,
one finds equivalently to (\ref{eq:e11}) but 
in terms of the Euclidean correlators \cite{CaronHuot:2009ns},
\begin{equation}
2 \int_{0}^{\infty} d \omega~ \frac{\delta \rho(\omega)}{\omega}
= \int_{0}^{\beta} d\tau~  [G(\tau) - G^{rec}(\tau)] ~.
\label{eq:ec4}
\end{equation}

\section{R-current correlator}
\label{sec:rcurrent}

\subsection{Strong coupling}

In \cite{Myers:2007we} the R-current correlator (at vanishing three-momentum)
in the ${\cal{N}} = 4 ~SYM$ theory is derived in the limit
of infinite 't Hooft coupling,  $\lambda \rightarrow \infty$.
The corresponding thermal spectral function reads
\begin{equation}
\rho^{strong}(\omega, T) = \frac{N_c^2 \omega^2}{32 \pi^2}
\frac{\sinh(\omega \beta/2)}{\cosh(\omega \beta/2) - \cos(\omega \beta/2)}~,
\label{eq:er1}
\end{equation}
and
\begin{equation}
\rho^{strong}(\omega, T=0) = \frac{N_c^2 \omega^2}{32 \pi^2}~.
\label{eq:er2}
\end{equation}

\noindent
From these expressions the integral (\ref{eq:e11})
 is obtained to vanish numerically, i.e. the sum rule in the 
strong coupling limit becomes,
\begin{equation}
2 \int_{0}^{+\infty} d\omega ~
\frac{\delta \rho^{strong}(\omega)}{\omega} = 0~.
\label{eq:er3}
\end{equation}

\noindent
This may be seen as follows: 
$\delta \rho^{strong}(\omega)$ is oscillating around
$\delta \rho^{strong}(\omega) = 0$
- as can be seen in Fig.~3b of \cite{Myers:2007we}. 
For large $\omega$ it reads
\begin{equation}
\delta \rho^{strong}(\omega) \simeq \frac{N_c^2 \omega^2}{16 \pi^2}
e^{- \omega \beta/2}~ \cos(\omega \beta/2)~,
\label{eq:er4}
\end{equation}
such that
\begin{equation}
\int_{0}^{\infty} dx ~x~ e^{-x} \cos x = \frac{\Gamma(2)}{2}
\cos(2 \arctan 1) = 0~.
\label{eq:er5}
\end{equation} 

\noindent
To make sure that for the sum rule (\ref{eq:er3})
$\delta G_R(\omega =0) - \delta G_R^{\infty} = 0$,
required by (\ref{eq:e11}),
let us use
\begin{equation}
G_R(\omega) = \Pi(\omega) = \Pi^L(\omega) = \Pi^T (\omega)~,
\label{eq:er6}
\end{equation}
for vanishing three-momentum and given in \cite{Myers:2007we}
by
\begin{equation}
\Pi(\omega) = \frac{N^2 T^2}{8}
\{ i \frac{\omega}{2 \pi T} + \frac{\omega^2}{(2 \pi T)^2}
[~ \psi(\frac{(1-i)\omega}{4 \pi T}) + \psi(-\frac{(1+i)\omega}{4 \pi T})~]\} ~,
\label{eq:er7}
\end{equation}
with  $\psi$ the logarithmic derivative of the gamma function.
This retarded correlator has a quasinormal spectrum
\cite{Berti:2009kk}
 with  poles located
at \cite{Myers:2007we}
\begin{equation}
\omega = 2 \pi T (\pm n - i n)~~, ~n = 1,2,.... 
\label{eq:er8}
\end{equation}

\noindent
From (\ref{eq:er6}, \ref{eq:er7}) follows
\begin{equation}
\delta G_R (\omega = 0) = 0~.
\label{eq:er9}
\end{equation}

\noindent
Using 
\begin{equation}
\psi(z) \rightarrow \ln{z} - \frac{1}{2 z} - \frac{1}{12 z^2}
+ O(\frac{1}{z^4})~,~~ z \rightarrow \infty~,
\label{eq:er10}
\end{equation}
it follows
\begin{equation}
\Pi(\omega = i \alpha) \rightarrow - \frac{N^2}{16 \pi^2}
\alpha^2 \ln{\alpha}~,
\label{eq:er11}
\end{equation}
in the limit $\alpha \rightarrow \infty$,
such that 
\begin{equation}
\delta G_R^{\infty} = 0~.
\label{eq:er12}
\end{equation}
This shows the validity of (\ref{eq:er3}).

\subsection{Weak coupling: free theory}

As in \cite{Teaney:2006nc} the spectral density of the free theory,
for  $\lambda =0$, denoted by $\rho^{L}_{JJ} (\omega)$ is used as 
a reference for comparison.
In Appendix A of \cite{Teaney:2006nc}  the one loop contribution
 for the ${\cal{N}} =4$ theory is
derived. 
Using this result the integral, 
\begin{eqnarray}
&& 2 \int_{0}^{+\infty} d\omega ~
\frac{\delta \rho^{weak}(\omega)}{\omega} \equiv
2 \int_{0}^{+\infty} d\omega ~
\frac{\delta \rho^{L}_{JJ}(\omega)}{\omega} 
\nonumber \\
&& = 2 \int_{0}^{+\infty} d\omega \{ \frac{N^2 T^2}{12} \delta (\omega) 
+ \frac{N^2}{48 \pi^2}~ \omega~ [~
n_B(\omega/2T) - 2 n_F(\omega/2T)~] \}
\nonumber \\
&& = \frac{N^2 T^2}{6}~,
\label{eq:w1}
\end{eqnarray}
is non-vanishing due to the contributions of the fermion and scalar
quasiparticles.

\section{Conclusion}

Indeed comparing (\ref{eq:er3}) with (\ref{eq:w1}) there is a significant
difference between the strong and weak coupling result for the sum rule
 of the current-current correlator, maybe even more significant
than the difference between the Euclidean correlators $G_{JJ}(\tau)$
for the interacting and free theory as shown in Fig.~3(b)
 in \cite{Teaney:2006nc}, which is at most $\sim 20 \% $.

\noindent
It is worth to keep in mind that these differences are coming from
the qualitative different analytic structures of the retarded
Green's function $G_R(\omega)$ as a function of $\omega$ in the 
strong and weak coupling limit, respectively \cite{Hartnoll:2005ju}.
For strong coupling  $G_R(\omega)$ (\ref{eq:er7}) contains only
poles, i.e. the infinite sequence of quasinormal modes (\ref{eq:er8})
obtained from
perturbations about the $AdS_5$ black brane space-time
\cite{Berti:2009kk,Son:2007vk}.
In the perturbative approximation of $G_R(\omega)$ quasiparticles
and branch cuts are present \cite{Petreczky:2005nh,LeBellac:1996}.

\noindent
Are QCD lattice calculations \cite{Aarts:2007wj,Aarts:2008zz}
able to show a similar significant difference between strong and weak coupling 
behaviour by considering the integrals (\ref{eq:ec4}) ?

\section*{Acknowledgements}

I would like to thank Paul Romatschke for useful comments. 


\end{document}